\documentclass[english,prl,twocolumn,showpacs,amsmath,amssymb,reprint]{revtex4-1}

\usepackage[english]{babel}
\usepackage{xcolor}
\usepackage{epsfig}
\usepackage{float}

\begin{document}
	
	\title{Morphological transition between patterns formed by threads of magnetic beads}
	\author{Danilo S. Borges$^{1}$, Hans J. Herrmann$^{1,2}$, Humberto A. Carmona${^1}$, Jos\'e S. Andrade Jr.$^{1}$ and Asc\^anio D. Ara\'ujo$^{1}$} 
	\affiliation{$^{1}$Departamento de F\'isica,
		Universidade Federal do Cear\'a, Campus do Pici,
		60455-760 Fortaleza, Cear\'a, Brazil\\
		$^{2}$PMMH, ESPCI, CNRS UMR 7636, 7 quai St. Bernard, 75005 Paris, France}

	\date{\today}
	
	\begin{abstract}
		Magnetic beads attract each other forming chains. We pushed such chains into an inclined Hele-Shaw cell and discovered that they spontaneously form self-similar patterns. Depending on the angle of inclination of the cell, two completely different situations emerge, namely, above the static friction angle the patterns resemble the stacking of a rope and below they look similar to a fortress from above. Moreover, locally the first pattern forms a square lattice, while the second pattern exhibits triangular symmetry. For both patterns the size distributions of enclosed areas follow power laws. We characterize the morphological transition between the two patterns experimentally and numerically and explain the change in polarization as a competition between friction-induced buckling and gravity. 
	\end{abstract}
	
	
	\maketitle

	The folding and crumpling of slender objects like wires is of increasing interest due to its many applications in mechanics and biology. A rich spectrum of instabilities and patterns has been found depending on friction, stiffness, aspect ratio and the type of confinement~\cite{stoop2008morphological,shaebani2017compaction,khadilkar2018self}. By adding attractive forces, self-assembling systems like origamis have been devised~\cite{shenoy2012self}. However, much less is known when the wire has a polarization, as it is the case for a chain of magnetic particles. In fact, threads of magnetic beads occur in Nature on different scales. They form on nanometric scale in magnetic colloids \cite{hill2014colloidal} and as chains of magnetosomes in magnetotactic bacteria \cite{hanzlik1996spatial,shcherbakov1997elastic,alphandery2012preparation}. Here we will consider macroscopic metal beads to study two-dimensional folding patterns by injecting them into a Hele-Shaw cell. The relative polarization of two chains allows for two fundamentally different types of local interactions which lead to new, completely dissimilar types of macroscopic patterns. In what follows, we show that the transition from one to the other can be controlled experimentally by adjusting the action angle of the gravitational forces on the system.

	\begin{figure}[!h]
		\centering
		\includegraphics[width=8.3cm]{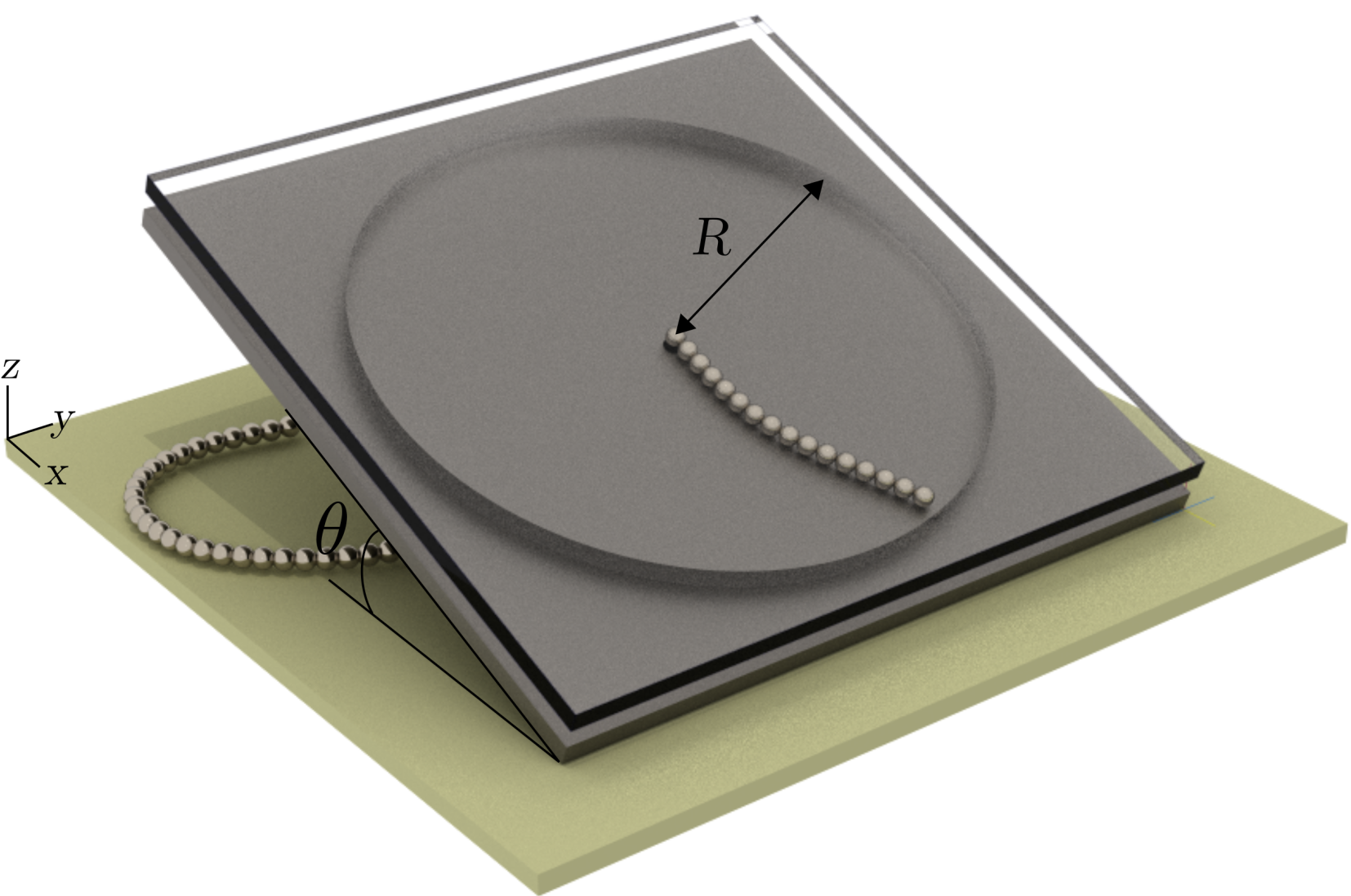}
		\caption{Schematic set-up of the experiment. The chain of magnetic beads is pushed from below into a Hele-Shaw cell with a transparent top plate inclined by $\theta $ degrees. The acrylic cylindric 
			ring of radius $R$ separates the two plates.}
		\label{fig:helle_shaw_inclination}
	\end{figure}

	The injection of wires into cavities has been of interest to model the coiling of long DNA in globules and viral capsids \cite{richards1973mode,kindt2001j,dai2016polymer,cao2017dna} as well as a minimally invasive treatment of saccular aneurysms \cite{johnston2002recommendations}. Fractal filling patterns have been observed, while the injection force diverges with a power law \cite{donato2002crumpled,donato2003scaling,gomes2010crumpled,lin2008crumpling}. Three different filling patterns emerge depending on friction and the bending elasto/plasticity of the wire: a spiral phase, a folding phase and a chaotic phase \cite{stoop2008morphological,vetter2013finite}. Also deformable cavities have been considered \cite{vetter2014morphogenesis,elettro2017drop}. In this work, we replace the wire by a chain of magnetic beads and the cavity by a Hele-Shaw cell. Magnetic hard spheres generate self-assembled patterns in the microscopic scale \cite{Mehdizadeh_Taheri14484,tripp2003flux,talapin2007dipole}, in the mesoscopic scale \cite{zahn2000dynamic,messina2015self,baraban2008frustration,klokkenburg2006quantitative} and in the macroscopic scale \cite{stambaugh2004segregation,vandewalle2014magnetic,stambaugh2003pattern}.
	The anisotropy of the magnetic forces  induces different orientations in the interaction between sections of the chain resulting in the self-assembly of novel types of patterns, which we realize here experimentally and numerically.

	The experiments were performed with magnetized neodymium beads of $d = 5 ~\text{mm}$  diameter. As shown in Fig.~\ref{fig:helle_shaw_inclination}, the Hele-Shaw cell consisted of a black plate covered with a transparent acrylic disk separated by a flat acrylic cylindric 
	ring of $6 ~\text{mm}$  height and radius $R$. The cell could be inclined by an angle $\theta$. A step motor controlled a quasi-static injection at $0.43 ~\text{mm/s}$  into a hole in the middle of the black bottom plate. 
	Images were recorded with a digital Canon PowerShot SX510 HS camera with $30$ frames per second at $27~\text{cm} $ above the cavity and used to determine the particle positions through image segmentation.

	\begin{figure*}
		\centering
		
		\includegraphics[width=\textwidth,height=5cm]{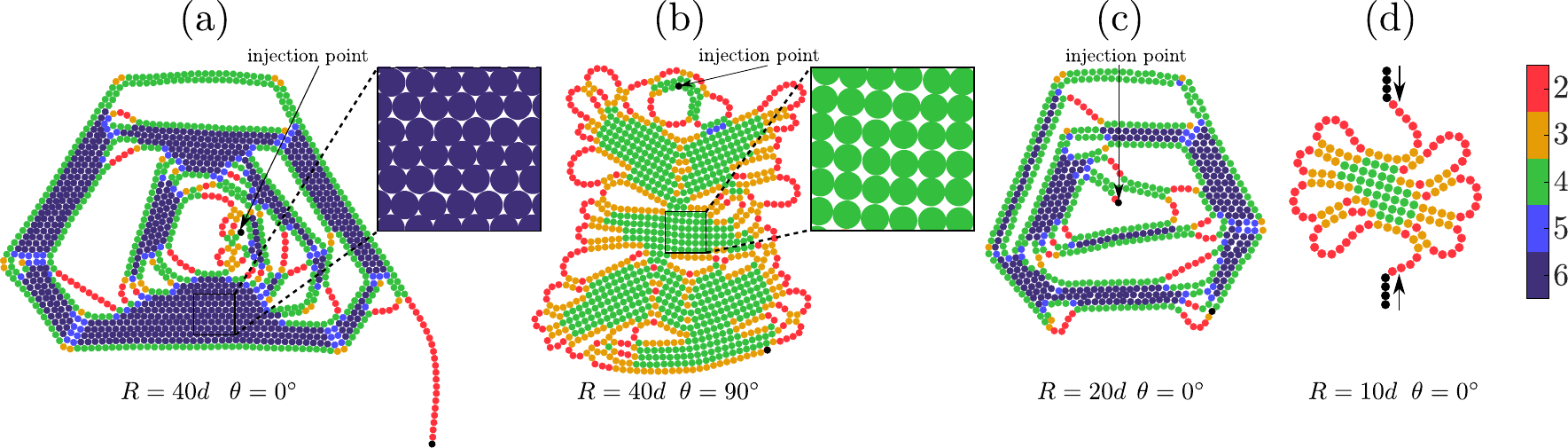}
		\caption{Patterns obtained after no more beads can be injected anymore into the cell for $\text{(a)}$, $R=40d$ and $\theta=0^{\circ}$; $\text{(b)}$, $R=40d$ and $\theta=90^{\circ}$; $\text{(c)}$, $R=20d$ and $\theta=0^{\circ}$. In $\text{(d)}$ two chains are injected into the cell from opposite sides, as indicated by black arrows, with $\theta=0^{\circ}$. For $\text{(a)}$ and $\text{(c)}$ we see \textit{polygonal} patterns and for $\text{(b)}$ and $\text{(d)}$ \textit{stacked} ones. The zooms in $\text{(a)}$ and $\text{(b)}$ highlight the two patterns, polygonal and stacked ones, respectively. The color code indicates the number of nearest neighbors of each particle, {\it i.e.}, having a distance of less than $10\%$ of their diameter. We see that in polygonal patterns locally most particles have six neighbors, producing thus a triangular lattice, while in stacked patterns four nearest neighbors prevail, yielding a square lattice. Only in $\text{(a)}$ the chains never hit the wall of the cell. The black points in $\text{(a)}$, $\text{(b)}$ and $\text{(c)}$ indicate the injection point.}
		\label{fig:patterns}
	\end{figure*}

	With aligned dipole moments, the magnetic beads naturally assemble into chains that exhibit macroscopically elastoplastic bending stiffness \cite{hall2013mechanics,vella2014magneto}. 
	We injected such chains of beads into the cavity from the bottom into the cell. When they enter the cell, the spheres in the chain and their magnetic moments undergo rotations, which weakens the forces in the direction of the chain between the beads close to the entrance. Eventually, either by hitting the wall of the cell or due to the friction with the bottom plate, the chain slows down and starts buckling. It then forms loops that get closer and closer to the injection hole, until the region close to the hole is jammed and no more particles can be inserted. At this point, the motor is stopped and the resulting pattern is analyzed. 
	
	In Figs.~\ref{fig:patterns} we show four of such patterns obtained for two different angles, $\theta=0^{\circ}$ and $90^{\circ}$, and cell radii, $R=20d$ and $40d$. For $\theta=0^\circ$ (see Figs.~\ref{fig:patterns}$\text{a}$ and \ref{fig:patterns}$\text{c}$), we observe the formation of {\it polygonal} shaped patterns which look like fortresses viewed from above. In the particular case of Fig.~\ref{fig:patterns}$\text{a}$, due to the larger size of the cell ($R=40d$), the chain never reaches the walls, but buckles before, stopping due to friction with the bottom plate. When the experiment is performed in a smaller cell ($R=20d$), as shown in Fig.~\ref{fig:patterns}$\text{c}$, the chain eventually hits the walls of the cell and then buckles, leading to the polygonal structure. In the case of $\theta=90^{\circ}$ (see Figs.~\ref{fig:patterns}$\text{b}$ and \ref{fig:patterns}$\text{d}$), the structure resembles the stacking of a rope, which we will thus call {\it stacked} patterns (see movies in the Supplementary Material \cite{Supplement}). 
	
\begin{figure}[!h]
		\centering
		\includegraphics[width=8.3cm]{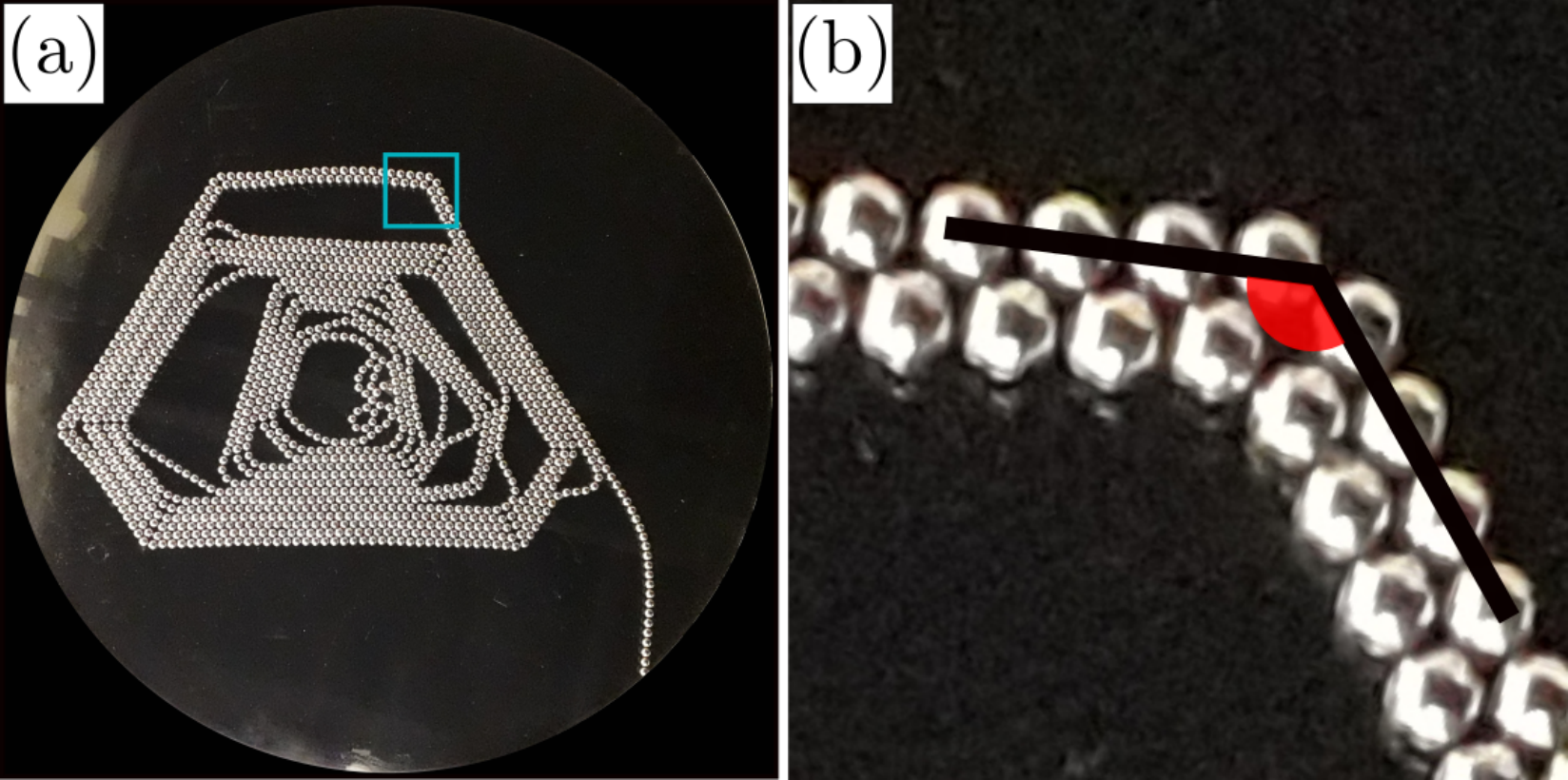}
		\caption{The \textit{polygonal} pattern is formed by stripes of paired chains with a locally triangular arrangement. These stripes spontaneously create corners forming angles of around $120^{\circ}$, as shown in $\text{(a)}$
			and enlarged in $\text{(b)}$.}
		\label{fig:angulo}
	\end{figure}

	In polygonal patterns, chain pieces having the same dipolar orientation attract each other, forming stripes that locally exhibit triangular symmetry. These stripes spontaneously bend, forming pronounced corners of around $120^{\circ}$, as shown in Fig.~\ref{fig:angulo}. In stacked patterns, on the other hand, chain pieces having opposite dipolar orientation attract each other forming domains that locally exhibit square symmetry (see schematic illustrations in Fig.~S1 of the Supplementary Material \cite{Supplement}).
	
	The reason why equally oriented chains form triangles, while chains of opposite orientation end up in squares, can be found by searching for the configurations of lowest magnetic energy $E$. This is obtained by summing over all pairs of beads in the pattern according to,
	\begin{equation}\label{eq.energy}
	E = -\sum_{i=1}^{N-1} \sum_{j = i+1}^{N} \pmb{m}_i\cdot \pmb{B}_j    ,
	\end{equation}
	
	\noindent where $\pmb{m}_i$ is the dipole moment of particle $i$, $N$ the total number of particles and $\pmb{B}_j$ the magnetic dipole field of particle $j$ at the position of particle $i$ given by, 
	
	\begin{equation}
	\label{eq:mag_field}
	\pmb{B}_j = \frac{\mu_0}{4\pi}\left[
	\frac{ 3\left(\pmb{m}_j\cdot \hat{\pmb{r}}_{ij}\right) \hat{\pmb{r}}_{ij}  
		- \pmb{m}_j }{r_{ij}^3} 
	\right],
	\end{equation}
	
	\noindent with $\pmb{r}_{ij}$ being the vector pointing from the center of particle $i$ to the center
	of particle $j$, $\pmb{\hat{r}}_{ij}=\pmb{r}_{ij}/|\pmb{r}_{ij}|$, and $\mu_0$ is the vacuum permeability. We performed Monte Carlo simulations of two parallel chains of magnetic dipoles using Eq.\eqref{eq.energy} as energy in the Boltzmann factor and found that, after decreasing the temperature, the energetically most favorable configurations for parallel (anti-parallel) orientations of the chains were indeed the triangular (square) configurations with all dipoles oriented in parallel along the direction of the chains. 
	
\begin{figure}[!h]
		\centering
		\includegraphics[width=8.3cm]{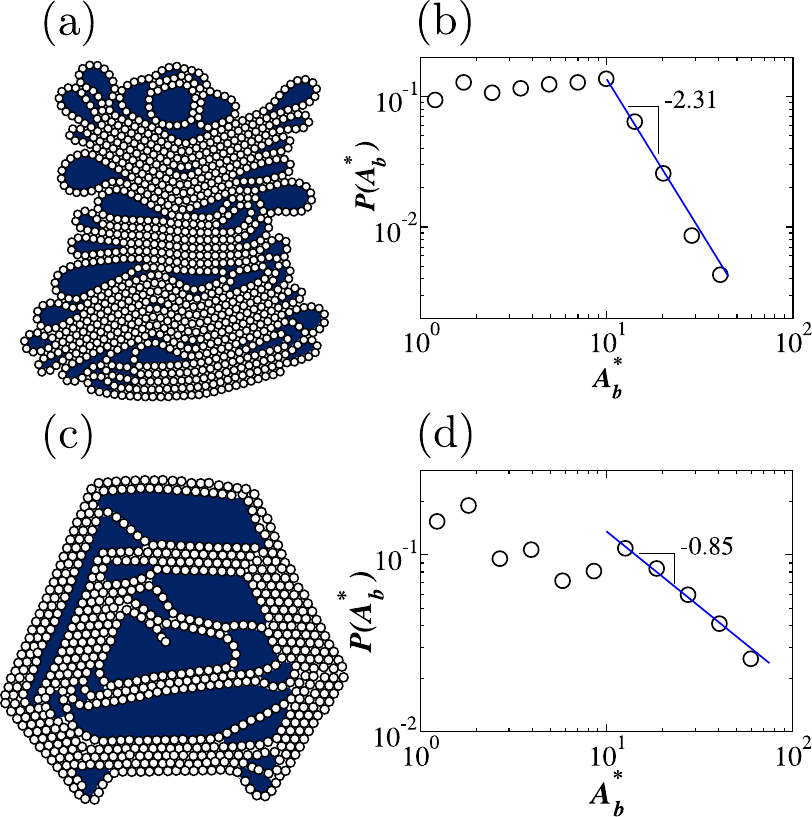}
		\caption{The areas enclosed by the loops within the chain of magnetic particles are shown in blue for two typical injection experiments resulting in $\text{(a)}$ a stacked pattern and $\text{(c)}$ a polygonal pattern. The dimensionless area $A^{\ast}_{b}=A_{b}/A_{0}$ is calculated from black-white images, where $A_{b}$ is the area of the hole measured in pixels and $A_{0}=\pi (d/2)^2$. The area distributions are shown in $\text{(b)}$ for the stacked and in $\text{(d)}$ for the polygonal patterns, calculated from the average over $15$ experimental realizations in each case. The blue solid lines are the least-squares fit to the data sets of a power law, $P(A^{*}_{b}) \propto {A^{*}_{b}}^{-\gamma}$, calculated for areas with more than $10$ pixels, with exponents $\gamma=2.31\pm 0.03$ and $0.85\pm 0.03$, for $\text{(b)}$ and $\text{(d)}$, respectively.}
		\label{fig:distribution}
	\end{figure}

	All patterns self-assemble into scale-free structures. For instance, this can be seen from the areas enclosed by loops, as shown in Figs.~\ref{fig:distribution}$\text{a}$ and \ref{fig:distribution}$\text{c}$. Clearly there are areas of all sizes. To calculate the areas, we first transform the RGB images into black-white images. The analysis is performed using the dimensionless area $A_b^{*}=A_b/A_0$, where $A_b$ is the area of the hole  and $A_0=\pi (d/2)^2$ is the area of the projection of one sphere, both measured in pixels. Typical examples of such two-dimensional disjointed domains are shown in Figs.~\ref{fig:distribution}$\text{a}$ and \ref{fig:distribution}$\text{c}$, obtained from the patterns in Figs.~\ref{fig:patterns}$\text{b}$ and \ref{fig:patterns}$\text{c}$, respectively. 
	The distributions of areas shown in Figs.~\ref{fig:distribution}$\text{b}$ and \ref{fig:distribution}$\text{d}$ 
	were obtained from the average over $15$ injection experiments each,  performed with inclination angles $\theta=0^{\circ}$ and $90^{\circ}$, respectively. As depicted, for areas that have at least $10$ pixels, both distributions follow power-law behaviors, $P(A^{*}_{b}) \propto {A^{*}_{b}}^{-\gamma}$, with exponents $\gamma=2.31 \pm 0.03$ for the stacked pattern ($\theta=0^{\circ}$) and $\gamma=0.85 \pm 0.03$ for the polygonal pattern ($\theta=90^{\circ}$).

\begin{figure}[h]
		\centering
		\includegraphics[width=8.3cm]{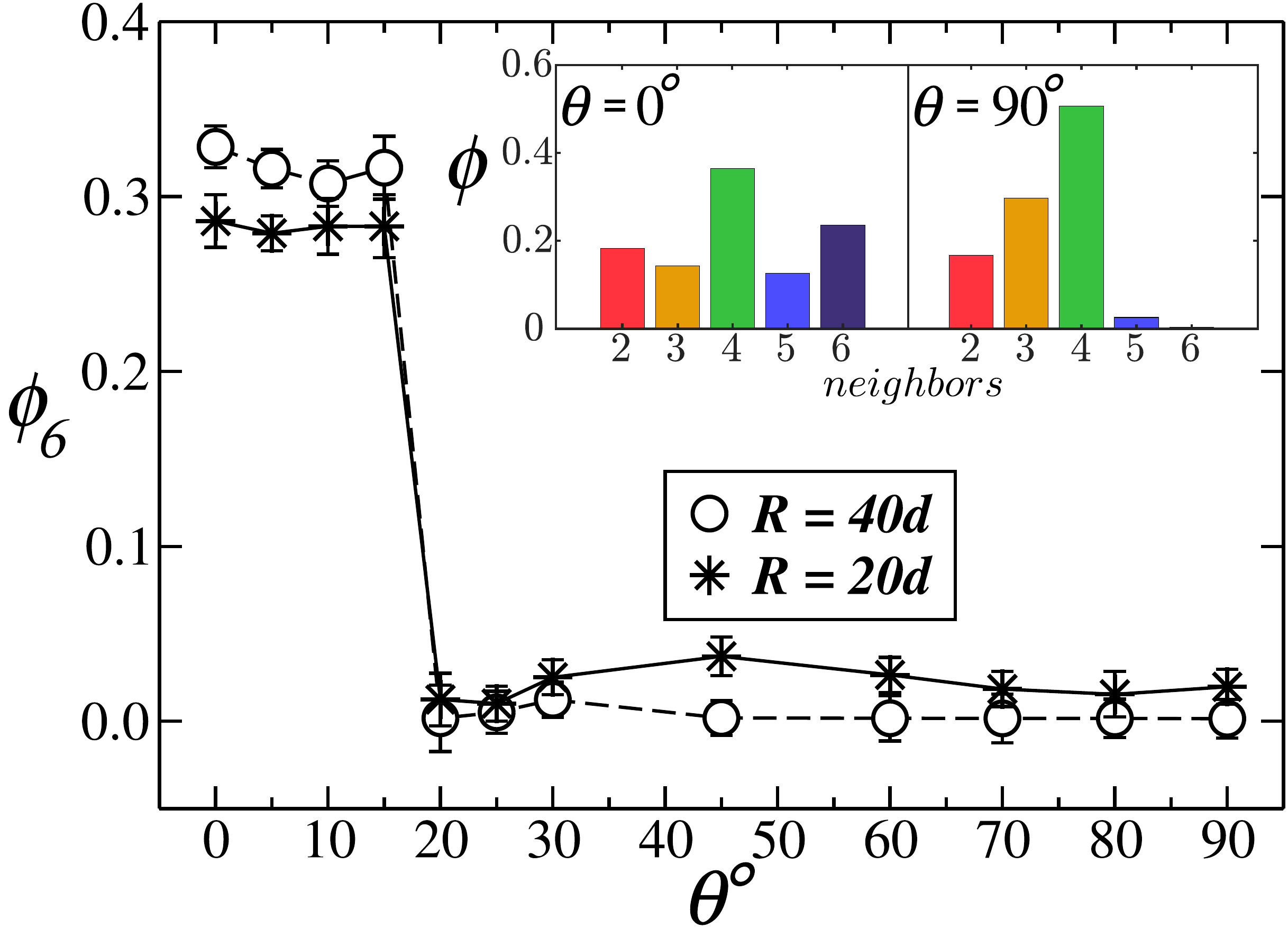}
		\caption{Order parameter $\phi_6$ of the morphological phase transition as a function of the inclination angle $\theta$ for $R=40d$ (circles) and $R=20d$ (stars). $\phi_6$ is defined as the average fraction of particles that have six neighbors. One clearly observes an abrupt jump near the critical point $\theta_{c}=19.4^{\circ}$, being the signature of a first-order transition. The inset shows how the average fractions $\phi$ of particles in the chain with $n$ neighbors change going from  $\theta=0^{\circ}$ to $\theta=90^{\circ}$, both for $R=40d$.}
		\label{fig:OP}
	\end{figure}

	By increasing the inclination angle $\theta$ of the cell we observe a transition from the polygonal to the stacked pattern. A convenient order parameter to characterize this transition is the percentage $\phi_6$ of particles that have six neighbors. In Fig.~\ref{fig:OP}, we plot $\phi_6$ as a function of $\theta$ for $R=40d$ and $20d$ and see that below a critical angle $\theta_{c}=19.4^{\circ}\pm 0.4$ the order parameter $\phi_6$ is finite and above it is zero. The change at $\theta_{c}$ is abrupt, as it is the case for first-order transitions, which is typically expected for morphological phase transitions. This sharp transition is also observed in the presence of finite-size effects, i.e., when the chain hits the wall of the cell for $R=20d$. The inset of Fig.~\ref{fig:OP} shows that the average fractions $\phi$ of particles in the chain having a certain number of neighbors $n$ ($2\leq n \leq 6$) change dramatically from $\theta=0^{\circ}$ to $\theta=90^{\circ}$.  
	
\begin{figure}[!h]
		\centering
		\includegraphics[width=8.3cm]{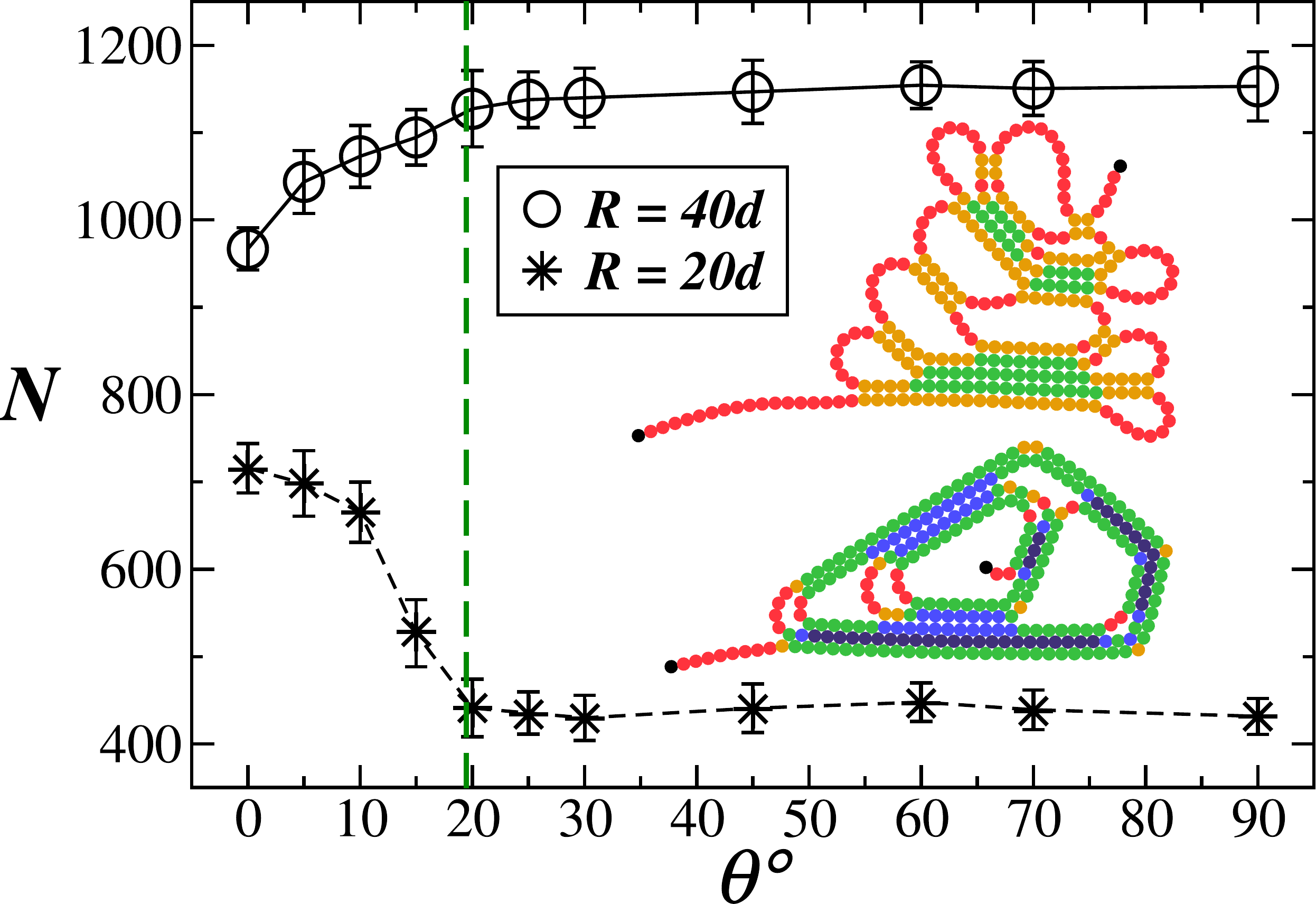}
		\caption{Number $N$ of particles in the pattern as a function of the inclination angle $\theta$ of the cell for $R=40d$ (full circles) and $R=20d$ (open circles). The vertical dashed green line at $\theta_{c}=19.4^{\circ}$ corresponds to the static friction angle of the particles with the bottom plate, measured by letting three particles forming a triangle slide down on the plate in order to avoid rolling. The relationship between the critical angle and the static friction coefficient is $\mu_e=\tan(\theta_{c})\approx 0.35$. Above $\theta_{c}$, the number of particles $N$ becomes independent of $\theta$, but depends on the radius $R$ of the cell if it is sufficiently small. In the inset we show simulated patterns for $\theta = 0^{\circ}$ without (upper) and with (bottom) weakening of the dipolar forces at the entry of the chain.}
		\label{fig:numero}
	\end{figure}

	In Fig.~\ref{fig:numero} we show how the number $N$ of particles that can be injected into the cell before the chain gets stuck depends on the inclination angle $\theta$. For $R=40d$, the size of polygonal patterns, i.e., below $\theta_{c}$, changes with $\theta$, while for stacked patterns, i.e., above $\theta_{c}$, the size is independent of $\theta$. If $R$ is too small ($R=20d$), the chain hits the wall of the cell and then the patterns can not attain their full size, with $N$ becoming substantially smaller, as depicted in Fig.~\ref{fig:numero}. 
	
	In order to measure the friction, we formed a triangle of three particles and let it slide down on the bottom plate. This triangular arrangement allowed to suppress any rolling. Interestingly, the inclination angle at which the triangle starts to slide down, i.e., the static friction angle, turns out to be $19.4^{\circ}\pm0.4$, agreeing perfectly with $\theta_{c}$. This seems to indicate that the phase transition is triggered by the sliding of the chain, that is, above (below) the static friction angle the chain will (not) slide therefore producing stacking (polygonal) patterns.
	
	Instead of pushing the chains of magnetic particles into the Hele-Shaw through a hole in the center, we also injected them into the cell from its boundary. Surprisingly, in this case it is impossible to create polygonal patterns, i.e., always only stacked patterns appear. An example for $\theta = 0^{\circ}$ in which two chains are simultaneously injected from opposite sides is shown in Fig.~\ref{fig:patterns}$\text{d}$.   
	
	
	In order to understand this last experimental observation and get deeper insight behind the mechanism producing the observed patterns, we also performed Discrete Element Model (DEM) simulations \cite{Cundall1979} using a fourth order Runge-Kutta algorithm for integration, and rotations \cite{Poschel2005}. The contact forces are written as, 
	
	\begin{equation}
	\vec{F}^c_{ij} = F_n \hat{n}_{ij}  + \vec{F}_s,
	\end{equation}
	where $F_{n}\hat{n}_{ij}$ and $\vec{F}_s$ represent the normal and shear forces between contacting spherical beads. The normal force is given by $F_n = k_n u_n$  where $k_n$ is the normal stiffness and $u_n$ is the contact overlap. The shear forces are computed incrementally by $\Delta \vec{F}_s = -k_s \Delta \vec{r}_{ij}$, where $k_s$ is the shear stiffness and $\Delta r_{ij}$ the relative shear displacement vector. Friction between particles is
	implemented in a similar fashion. The magnetic force $\vec F_{ij}^M$ and torque $\vec \tau_{ij}^M$ between particles $i$ and $j$ are assumed to be due to point-like magnetic dipoles at the
	centers of the beads, and computed as
	\begin{equation}
	\vec{F}^M_{ij} = \nabla \left(\vec{m}_i\cdot\vec{B}_j\right),
	\end{equation}
	and 
	\begin{equation}
	\vec{\tau}^M_{i,j} = \vec{m}_i \times \vec{B}_j,
	\end{equation}
	\noindent where $\vec{m}_i$ is the dipole moment of particle $i$ and $\vec{B}_j$ the magnetic dipole field of particle $j$ at the position of particle $i$, as defined in Eq.~\eqref{eq:mag_field}. The magnitude of the magnetic dipoles defines the bending stiffness of the chain.
	
	In our simulation, we inserted chains of magnetic particles quasi-statically either from the center or from one point at the boundary of the cell and always stacked patterns were formed independently of the choice of parameters, as shown in Fig.~\ref{fig:numero}~(upper inset). Only after weakening systematically the magnetic dipole force between the last two beads that just entered the cell by at least a factor of two, we could reproduce the polygonal pattern, as shown in Fig.~\ref{fig:numero}~(bottom inset). In fact, when particles are injected through a hole in the center of the cell, a rotation of $90^{\circ}$ is imposed on them, which locally weakens their magnetic forces. This weakening has a dramatic consequence on the evolution of the pattern (see the movies at the Supplementary Material \cite{Supplement}). After the chain has stopped either due to friction or after hitting a wall, it buckles in one direction. If no weakening is imposed at the entry, after some time the bending stiffness of the chain will however force it to flip back inducing an oscillatory behavior and forming stacking patterns. Only if at the entry the dipoles are sufficiently weakened, the chain can bend enough to allow it to continue turning in the same direction, forming, for adequately chosen parameters, the polygonal patterns.
	
	Here we reported new patterns that appear while feeding a chain of magnetic beads into the center of a 
Hele-Shaw cell. We discovered that, depending on the inclination of the cell, i.e., the effect of gravity, two completely different types of scale-free patterns self-assemble. The first-order phase transition between the two patterns occurs at the static friction angle. Crucial for obtaining the polygonal pattern is the weakening of the dipolar forces at the entry into the cell, due to the redirection of the chain by $90^{\circ}$.
The subtle effects encountered at the injection point discovered here, which dictates the way the magnetic chain deforms into the cell, might become relevant in the manipulation of magnetic colloids \cite{philipse2002magnetic}, chains of magnetosomes \cite{zhu2016measuring,katzmann2011magnetosome} and soft micromachines based on dipole-dipole interactions \cite{huang2016soft}. It would be interesting to include in the future electric, entropic, van der Waals and other forces. It would also be important to study patterns formed  by interacting chains filling three-dimensional cavities and investigate the effect of the magnetic dipoles and friction on those patterns.

	\begin{acknowledgments}
		We  thank  the  Brazilian  agencies  CNPq,  CAPES and  FUNCAP, and the  National Institute  of  Science  and  Technology  for  Complex  Systems (INCT-SC) in Brazil for financial support.
	\end{acknowledgments}
	
	
	
%
	
\end{document}